\documentstyle[aps,prd,twocolumn,epsfig]{revtex}
\def\b{\bar}
\def\d{\partial}

\def\cD{{\cal D}}

\def\m{\mu}
\def\n{\nu}

\def\t{\tau}

\def\~{\tilde}

\def\bY3{\bar Y_{,3}}
\def\Y3{Y_{,3}}
\def\z{\zeta}
\def\Z{{\b\zeta}}
\def\Y{{\bar Y}}
\def\cZ{{\bar Z}}
\def\`{\dot}
\def\be{\begin{equation}}
\def\ee{\end{equation}}
\def\bea{\begin{eqnarray}}
\def\eea{\end{eqnarray}}

\def\cF{{\cal F}}

\def\mn{{\mu\nu}}

\begin{document}
\title{Rotating ``Black Holes''
 with Holes in the Horizon}

\author{Alexander Burinskii\\
Gravity Research Group, NSI, Russian Academy of
Sciences, \\
B. Tulskaya 52  Moscow 115191 Russia;}

\author{Emilio Elizalde\\
Consejo Superior de Investigaciones Cient\'{\i}ficas (ICE/CSIC)
\\ \&
Institut d'Estudis Espacials de Catalunya (IEEC)\\
Campus UAB, Facultat de Ci\`encies \\ Torre C5-Parell-2a planta,
E-08193 Bellaterra (Barcelona), Spain;}

\author{Sergi R. Hildebrandt\\
Instituto de Astrofisica de Canarias,\\
C/Via Lactea s/n, La Laguna, Tenerife, 38200, Spain;}

\author{Giulio Magli \\
Dipartimento di Matematica del Politecnico di Milano,\\
Piazza Leonardo Da Vinci 32, 20133 Milano, Italy.}
\maketitle

\begin{abstract}
 Kerr-Schild solutions of the Einstein-Maxwell field equations,
containing semi-infinite axial singular lines, are investigated.
  It is shown that axial singularities break up the black hole,
forming holes in the horizon. As a result,
 a tube-like region appears which allows matter to escape from the
interior without crossing the horizon.
It is argued that axial singularities of this kind, leading to
very narrow beams, can be created
in black holes by external electromagnetic or gravitational
excitations and may be at the origin of astrophysically observable
effects such as jet formation.
\end{abstract}

\bigskip

{\bf 1. Introduction.} Most of the applications of General
Relativity to the astrophysics of compact objects are based on the
Kerr solution of the Einstein equations, which describes the
rotating stationary black hole. Of course, although the Kerr
solution brilliantly describes stationary phases of these objects,
it can hardly trace nonstationary behaviors, like bursts and jet
formation. However, the Kerr metric is only one member of a wider
class of Einstein-Maxwell solutions, the Kerr-Schild class. Here
we will consider the rotating Kerr-Schild solutions containing
axial semi-infinite singular lines. To the best of our knowledge,
these solutions have, so far, not received enough attention in
astrophysics, and it seems that they have never been analyzed in
detail from the physical point of view.

One of the first mentions of axial singular lines in the exact
solutions of the  Einstein-Maxwell field equations can be found in
the paper by Robinson and  Trautman \cite{RobTra}. Axial
singularities, which are analogs of the Dirac monopole, appear in
the NUT and Kerr-NUT solutions \cite{KraSte} and are related to
magnetic charge and to some ``magnetic type'' of mass which does
not have, so far, a clear (astro)physical interpretation, at least
for isolated sources. In the work of Debney, Kerr and Schild
\cite{DKS}, a broad class of stationary solutions of the
Einstein-Maxwell field equations was considered, among which
are the solutions containing axial semi-infinite singular lines,
to be investigated in what follows. Our aim is, in particular, to
investigate in detail the structure of horizons for these
solutions, to discuss some of the effects which the appearance of
the axial singularities may cause in the rotating astrophysical
sources, and to consider the physical consequences which may
originate from these singularities in initially stable black holes.
We show that axial singularities ``break up"  the Kerr black hole,
forming a hole in the horizon which connects the internal with the
external regions. As a result, although a ``horizon'' is still
present, it does not isolate
 the Kerr singularity from the exterior any more, and it turns out to be
``half dressed''. These results add this class of solutions to the
(rather restricted) family of solutions which have a clear physical
meaning, and, therefore, we believe they are of interest {\it per se}.
Moreover, we conjecture here that the formation of axial
singularities may result in the production of jets \cite{jets},
thus providing an alternative model for important astrophysical
phenomena.\medskip

{\bf 2. Structure of the Kerr solution.}-- The Kerr-Newman metric in
Kerr -Schild form can be written as \be g^{\mn}=\eta ^{\mn} - 2h
k^\m k^\n , \label{ksa}\ee where $k^\m$ is a null vector field $k_\m
k^\m =0$ which is tangent to the Kerr principal null congruence. The
function $h$ has the form $h= \frac {Mr-e^2/2} {r^2 + a^2 \cos^2
\theta},$ where the oblate coordinates $r,\theta$ are used on the
flat Minkowski background $\eta^{\mn}$. In the case of rotating Kerr
solutions, the Schwarzschild horizon splits into four surfaces: two
surfaces of the staticity limit, $ r_{s+}$ and $r_{s-}$, which are
determined by the condition $g_{00} =0$, and two surfaces of the
event horizons, $S(x^\m)=$ const, which are the null surfaces
determined by the condition $ g^\mn (\partial _\m S) (\partial _\m
S) =0. $ In the case $e^2 +a^2
>M^2$, the horizons of the Kerr-Newman solution disappear and the
Kerr singular ring turns out to be naked. To avoid the problems
with a two-sheet topology, this singularity may be covered by a
smooth disklike source \cite{behm,BurBag}.
\medskip

{\bf 3. Rotating solutions with axial singularities.} -- The
Kerr metric above is actually only a special case of a much
more general class of exact solutions of the Einstein-Maxwell
field equations with axial singularities, discovered by
Debney, Kerr and Schild in the seminal paper \cite{DKS}. These
solutions are obtained considering the Kerr-Schild ansatz
$g^{\mn}=\eta ^{\mn} - 2h k^\m k^\n ,$ with a null vector field
$k^\m$ ($k_\m k^\m =0$), which is tangent to a geodesic and
shear-free principal null congruence (PNC). As a result, the
spacetimes are algebraically special, and many tetrad Ricci
components vanish, which leads to a strong  restriction on the
tetrad components of the electromagnetic field as well.
We skip here the details of the derivation, to mention only the fact
that the corresponding solutions turn out to be aligned with the
Kerr PNC, in the sense that \be F^\mn k_\m=0 \label{align}.\ee The
Debney-Kerr-Schild solutions arise when one more restriction on
the electromagnetic field is imposed, which leads to
stationary solutions without electromagnetic waves. The
unique non-zero component of the field tensor in this case is $\cF
_{31} = -(AZ),_1$ where $Z$ is the (complex) expansion of the PNC.
The function $A$ has the general form \be A= \psi(Y)/P^2, \label{AY} \ee
where $P=2^{-1/2}(1+Y\Y)$, and $\psi$ is an arbitrary holomorphic
function of $Y$. The resulting metric has the Kerr-Schild form
(\ref{ksa}), where the function $h$ is given by \cite{DKS} \be
h=M(Z+\bar Z)/(2P^3) - A \bar A Z\bar Z /2. \ee
In terms of spherical coordinates on the flat background one has
$Y(x) = e^{i\phi} \tan \frac {\theta} 2$, which is singular at
$\theta=\pi$. This singularity will be present in any
holomorphic function $\psi (Y)$, and, consequently, in $A$ and in
$h$. Therefore, all the solutions of this class ---with the
exclusion of the case $\psi =$ const which corresponds to the
Kerr-Newman solution--- will be singular at some angular direction
$\theta$.

The simplest cases are $\psi=q/(Y +c)$ and $\psi=q(Y +b)/(Y+c)$,
which correspond to an arbitrary direction of the axial
singularity. However, the sum of singularities in different
directions is also admissible $\psi(y)=\sum_i q_i(Y
+b_i)/(Y+c_i)$, as well as polynomials of higher degree.

We should stress here that the above solutions do not
contradict the assertion of the ``no hair theorem" of Carter and
Robinson, which states uniqueness of the Kerr and Kerr-Newman
solutions \cite{Cha} under certain {\it regularity} hypotheses,
since these
are here {\it not} satisfied here. Indeed, the spacetime has to be
asymptotically flat and regular out of the horizon, conditions
that are not fulfilled, since the axial singularity extends to
infinity. Also, the assumptions on the structure of the horizon
are not fulfilled. Indeed, the horizon of a black hole should be
smooth and convex. But here the horizon is not smooth at the
bifurcation points, and does not form a convex surface. In this
connection it may be noted that, recently,  multiparticle
Kerr-Schild solutions \cite{Multiks} have been obtained. In the
case of the two-particle solution, the Kerr-Newman solution
exhibits two semi-infinite axial singularities which are caused by
interaction with an external particle and are oriented along the
line connecting the two particles. \medskip

{\bf 4. Structure of the horizons and diagrams of the maximal analytic
extension (MAE).} --
Let us  now
analyze the structure of the horizons for the solutions
containing axial singular lines.
It is convenient to use the metric in the Kerr
coordinates $(t,r,\theta,\phi)$, \cite{DKS}.
The metric can be represented in the form
\bea  ds^2=(2H-1) dt^2 +
\Sigma (d\theta^2 + \sin ^2\theta d \phi ^2) + \\ \nonumber
2 (dr -a\sin ^2\theta d\phi)^2 +
(2H-1)(dr-a \sin^2\theta d\phi)^2 - \\ \nonumber
4H dt (dr -a \sin^2\theta d\phi) , \label{ksm}
\eea
where $\Sigma=r^2+a^2\cos ^2 \theta,$  $H =Mr/\Sigma$ for
vacuum metric and
$H =(Mr- e^2/2)/\Sigma$ for the charged Kerr-Newman solution.

The simplest axial singularity is
the pole $\psi =q/Y.$ In this case the
function $H$ takes the form
\be H= [Mr - \frac 12 (q/\tan \frac \theta 2)^2  ]/\Sigma
\ee
The  boundaries of the
ergosphere, $r_{s+}$ and $r_{s-}$ (which are determined by the
condition $g_{00} =0$ or $2H-1=0$), are the solutions of the
algebraic equation \be 2Mr - r^2 -(q/\tan \frac \theta 2)^2 - a^2
\cos ^2 \theta =0. \label{g00h} \ee This solution acquires a new
feature: the surfaces $r_{s+}$ and $r_{s-}$ turn out to be joined
by a tube, forming a simply connected surface (see Figs. 1-3)
which has the topology of a sphere.

The surfaces for the event horizons are null and obey the
differential equation  \be (\d_r S)^2 [ r^2 +a^2 +(q/\tan \frac
\theta 2)^2 -2Mr ] - (\d_{\theta} S)^2 =0 \label{horh}. \ee One can
ignore the time and $\phi$ dependencies of the surface $S$. However,
the coefficients do depend on  $\theta$, and $(\d_{\theta} S)^2\ne
0,$ which does not allow us to get an explicit solution.
 Fortunately, $(\d_{\theta} S)^2$ is very small for
most of the horizon surface, with the exclusion of a small
$\theta-$vicinity of the axial singularity. We represent the surface
$S$ in the form $r=r(\theta)$ and
obtain an approximate solution which is very close to the exact
one for the regions where $(\d_{\theta} r)^2 $ is small.
Neglecting the term  $(\d_{\theta} r)^2$, one obtains that an
approximate form of this surface is described by the equation $
r^2 +a^2 +(q/\tan \frac \theta 2)^2 -2Mr =0 . $
The resulting surfaces are shown in Figs.~1-3 for different values
of the parameters $M$, $a$ and $q$. The axis of symmetry is the
horizontal z -axis, and the axial singular beam is directed along
the direction of angular momentum.

\begin{figure}[ht]
\centerline{\epsfig{figure=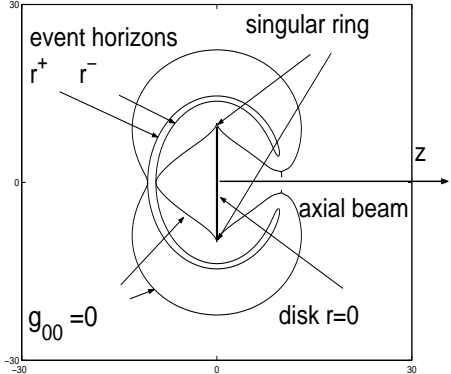,height=5.1cm,width=6cm}}
\caption{{\protect\small
Near extremal black hole with a hole in the horizon, for
$M=10, \  a=9.98, \ q=0.1$. Represented is the
axial section. The singular ring is placed at the boundary of the
disk $r=0.$ The event horizon is a closed connected surface
surrounded by the closed connected surface $g_{00}=0$. }} \end{figure}

\begin{figure}[ht]
\centerline{\epsfig{figure=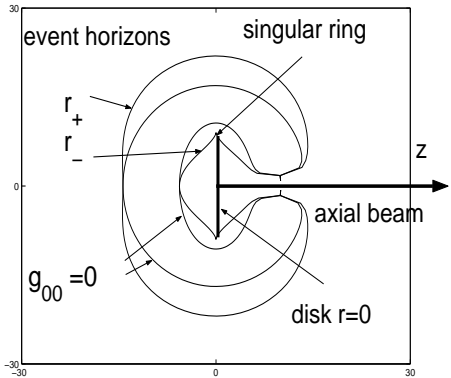,height=5.1cm,width=6cm}}
\caption{{\protect\small  Large hole in the horizon, for $M=10, \  a=9, \ q=0.3$.
Axial section. The singular ring is placed at the boundary of the
disk $r=0.$
The event horizon is a closed connected surface surrounded by the
closed connected surface $g_{00}=0$. }}
\end{figure}

\begin{figure}[ht]
\centerline{\epsfig{figure=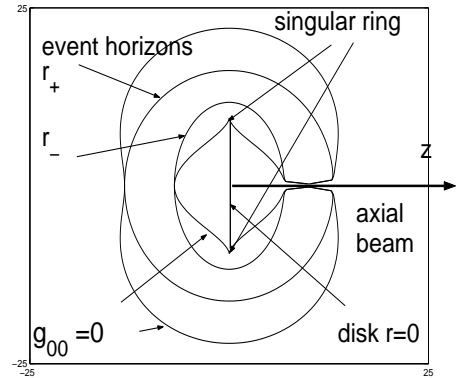,height=5.1cm,width=6cm}}
\caption{{\protect\small Narrow hole in the horizon, for $M=10, \  a=9.5, \ q=0.03$.
Axial section. The singular ring is placed at the boundary of the
disk $r=0.$
The event horizon is surrounded by the closed surface of boundary of
the ergosphere, $g_{00}=0$. }}
\end{figure}

Note that, for physical reasons,
the event horizon has to lie {\it inside} the boundaries of the
ergosphere. We clearly see that the resulting surfaces satisfy this
requirement.
 Similar to the boundary of the ergosphere, the two event
horizons are joined into one connected surface with spherical
topology, and the surface of the event horizon lies inside the
boundary of the ergosphere. As it is seen from the figures, the
axial singularities lead to the formation of the holes in the black
hole horizon, which opens up the interior of the ``black hole'' to
external space.

The structure of the diagrams of the maximal analytic extension
(MAE) \cite{Car} in these solutions depends on the section
considered. If the section is chosen away from the axial
singularity and  corresponding tube-like region, the diagram of
the MAE will be just the same as for the usual solution for a
rotating black hole. If the section goes through the axial
singularity, the tube-like hole in the horizon leaves a trace on
all patches of the MAE. The $r_+$ and $r_-$ surfaces are deformed
and approach towards each other, joining at some distance from
the axial
singularity and forming an access duct to the interior of the
former black hole. Therefore, tube-like channels connecting the
interior and the exterior at some angular direction will appear on
all patches of the diagram.

These {\it black holes with holes in the horizon} have thus
{\it preferred directions} along which the causal structure differs from
that of `true' black holes. Their singularity is, therefore,
naked, but the nakedness is of a very peculiar type, since it
manifests itself in specific directions only. A similar situation
occurs with other non-spherical exact solutions, like e.g. the so
called Gamma metric \cite{hmm}.
\medskip

{\bf 5. Possible astrophysical applications.} -- Two main questions
appear in confronting astrophysical applications of these
solutions: (i) which consequences will follow from the existence
of the holes in the black hole horizon, and (ii) which kind of
mechanism could lead to the appearance of the axial singularity
and corresponding holes in the horizon?

Note that axial singularities may carry travelling electromagnetic
and gravitational waves which propagate along them as along
waveguides, a phenomenon described by exact singular pp-wave
solutions of the Einstein-Maxwell field equations \cite{KraSte}.
The appearance of the axial singularities in rotating
astrophysical sources may be related to their {\it excitations} by
gravitational and/or electromagnetic waves, and has to be
necessarily caused by some nonstationary process. It was argued in
\cite{BurTwi,BurAxi} that electromagnetic excitation of black
holes leads inevitably to the appearance of axial singularities.
The motivation for this statement is based on the treatment of the
exact aligned wave solutions of the Maxwell equations on the Kerr
background, satisfying (\ref{align}), since only the  aligned wave
solutions are consistent  with the geodesic and shear-free
principal null congruence  of the Kerr geometry, and only those
may be used as candidates for the exact consistent solutions of
the Einstein-Maxwell field equations.

Similar to the stationary case considered in \cite{DKS}, the general
aligned solutions  are described by two self-dual tetrad components
$\cF _{12} = AZ^2$ and $\cF _{31} = \gamma Z -(AZ),_1$,
 where the function $A$ acquires an extra parameter
$\t$ which is a complex retarded-time \cite{BurAxi} (see the
appendix for details). The simplest wave modes \be\psi _n = q Y^n
\exp {i\omega _n \t} \equiv q (\tan \frac \theta 2)^n \exp {i(n\phi
+ \omega _n \t)} \ee can be labeled by the index $n=\pm 1, \pm 2,
...$, which corresponds to the winding number for the phase wrapped
around the axial singularity. The leading wave terms have the form $
\cF |_{wave} =f_R \ d \z \wedge d u  + f_L \ d \Z \wedge d v ,$
where $f_R = (AZ),_1$ and $f_L =2Y \psi (Z/P)^2 + Y^2 (AZ),_1$ are
the factors describing the ``left'' and ``right" waves propagating,
correspondingly, along the $z^-$ and $z^+$ semi-axes.
 Near the $z^+$ axis,  $|Y|\to 0$,
and for $r \to \infty $, we have $Y \simeq  e^{i\phi} \frac \rho
{2r}$, where $\rho$ is the distance from the $z^+$ axis.
Similarly, near the $z^-$ axis $Y \simeq  e^{i\phi} \frac {2r}
\rho  $ and $|Y|\to \infty$. For $|n|>1$ the solutions contain
axial singularities which do not fall off  asymptotically, but are
increasing, denoting instability. The leading wave for $n=-1$, \be
\cF^+_{-1}= - \frac {4q e^{-i2\phi+i\omega _{-1} \t_+ }} {\rho ^2}
\ d \z \wedge du , \ee is singular at the $z^+$ semi-axis and
propagates to $z=+\infty,$  while for $n=1$, \be \cF^-_{1}= \frac
{4q e^{i2\phi+i\omega _{1} \t_- }} {\rho ^2} \ d \Z \wedge dv ,
\ee the singularity is at the $z^-$ semi-axis and the wave propagates
to $z=-\infty .$

By considering the limiting fields near the singular axis, one can
find the corresponding self-consistent solutions of the
Einstein-Maxwell field equations \cite{BurAxi}. They are singular
$pp-$waves \cite{KraSte} having the Kerr-Schild form of the metric
(\ref{ksa}) with a constant vector $k^\m$.  In particular, the
wave propagating along the $z^+$ axis has $k_\m dx^\m= - 2^{1/2}du
$. Therefore, wave excitations of the Kerr geometry lead to the
appearance of singular $pp-$waves which propagate outward along the
axial singularities. In real situations, axial singularities
cannot be stable and they will presumably correspond to some type
of jet or burst, hence it is natural to conjecture that the related
holes in the horizon will also be at the origin of jet
formation.

Observational evidence shows a preference for two-jetlike sources,
as e.g., in the field of radio loud sources
\cite{punsly2001,antonnuci1993}.
These jets are emitted in opposite directions along the same axis.
This scenario corresponds to our treatment
of the sum of two singular modes with $n=\pm 1$ and
to two oppositely positioned holes in the horizon. The results are
basically the ones obtained for the
single beam case, applied to $ 0 \le \theta \le \pi/2 $ and
$ \pi/2 \le \theta \le \pi $.

It is known that the Kerr solution has a repulsive gravitational
force acting on the axis of symmetry for $r<a$. It can be described
by the Newton potential $\Phi(r,\theta)= -2h = -2Mr/(r^2+a^2).$ In
the small vicinity of the axial singularity, a
gravitational repulsion from the singularity will appear too,
since the potential
acquires the form \be \Phi(r,\theta)= -2h = (-2Mr
+|\psi|^2)/(r^2+a^2), \ee where $|\psi|^2 \sim q^2 \tan ^2 \frac
\theta 2.$ Quantum processes of pair creation near the axial
singularity will also take place, and presumably be responsible
for its regularization \cite{BirDev}.

On the other hand, electromagnetic $pp-$waves along the singularity
will cause a strong longitudinal pressure pointed outwards from the
hole. It can be easily estimated for the modes of the $pp-$waves with
$n=\pm 1$ \cite{BurAxi}.
For example, the corresponding energy-momentum tensor
is $ T^\mn = \frac 1 {8\pi} |\cF^+_{-1}|^2 k^\m k^n ,$ and the wave
beam with mode $n=-1$, propagating along the $z^+$ half-axis, will
exert
the pressure \be p_{z^+} = \frac {2q^2 e^{ 2 a \omega _{-1}}} {\pi
\rho ^4} , \ee where $\rho$ is an axial distance from the
singularity and $\omega _{-1}$ the frequency of this mode. For the
exact stationary Kerr-Schild solutions, one can use this expression
in the limit $\omega_{-1}=0$. \medskip

{\bf Conclusions.}-- From the analysis above, we conclude that the
aligned excitations of the rotating black hole (or naked rotating
source) lead, unavoidably, to the appearance of axial singularities
accompanied by outgoing traveling waves and also to the formation of
holes at the horizon, which can lead on its turn to the production
of astrophysical jets \cite{jets}.

Multiparticle Kerr-Schild solutions \cite{Multiks}
suggest that axial singularities are to
 be bi-directional and oriented along the line connecting
the interacting particles. Thus, it will be interesting to analyze
in further detail the observed jets in order to check the conjecture
that they may be indeed
triggered by radiation coming from  remote active objects.

Finally, one may suspect this effect to be related to the known
phenomenon of superradiance, although the usual treatment of the
latter does not take into account the condition (\ref{align}),
which specifically leads to the formation of narrow beams.
\medskip

{\bf Acknowledgments.} AB has been supported by the RFBR project
04-0217015-a. EE has been supported by DGICYT (Spain), project
BFM2003-00620 and SEEU grant PR2004-0126, and by AGAUR (Generalitat
de Catalu\-nya), contract 2005SGR-00790. \bigskip

{\bf Appendix: Aligned e.m. solutions on the Kerr-Schild
background.}-- The aligned field equations for the Einstein-Maxwell
system in the Kerr-Schild class were obtained in \cite{DKS}. The
electromagnetic field is given by tetrad components of the self-dual
tensor $\cF_{12} =AZ^2 , \quad \cF _{31} =\gamma Z - (AZ),_1   , $
where commas denote the directional derivatives with respect to
the chosen null
tetrad vectors. The equations for the electromagnetic field are \be A,_2 - 2
Z^{-1} \cZ Y,_3 A  = 0 , \label{3}\ee \be \cD A+  \cZ ^{-1} \gamma
,_2 - Z^{-1} Y,_3 \gamma =0 . \label{4}\ee
%Gravitational field equations
%yield \be M,_2 - 3 Z^{-1} \cZ Y,_3 M  = A\bar\gamma \cZ ,
%\quad \cD M  = \frac 12 \gamma\bar\gamma  ,
%\label{6}\ee
where $ \cD=\d _3 - Z^{-1} Y,_3 \d_1 - \cZ ^{-1} \Y ,_3 \d_2   \ . $
Solutions of this system were given in \cite{DKS} only for the
stationary case for $\gamma=0$, while the oscillating e.m. solutions
correspond to the case $\gamma \ne 0$.

For the sake of simplicity we consider the gravitational
Kerr-Schild field as stationary, although in the resulting e.m.
solutions the axial symmetry is broken, which has to lead to
oscillating backgrounds if the back reaction will be taken into
account.
%The recent progress in the obtaining the nonstationary solutions
%of the Kerr-Schild class is connected with introduction of a
Define a complex retarded time parameter $\t = t_0 +i\sigma = \t
|_L$ \cite{Bur-nst},
%which
%is determined as a result of the intersection of the left (L) null
%plane and the complex world line \cite{Bur-nst}. The left null
%planes are the left generators of the complex null cones and play
%a role of the null cones in the complex retarded-time
%construction.
which satisfies the relations \be (\t),_2 =(\t),_4 = 0 \ . \label{7}
\ee
%It allows one to represent
%the
Eq.~(\ref{3}) becomes $(AP^2),_2=0$,
%in the form \be (AP^2),_2=0 \ , \label{8}
%\ee
which can be integrated, yielding $ A=\psi(Y,\t)/P^2 $. It has the
form obtained in \cite{DKS}. The only difference is in the extra
dependence of the function $\psi$ from the retarded-time parameter
$\t$. This means, that the stationary solutions obtained in \cite{DKS}
may be considered as a low-frequency limit of these solutions.

One can easily check that the action of the operator $\cD$ on the
variables $Y, \bar Y $ and $ \rho$ is  \be \cD Y = \cD \bar Y =
0,\qquad \cD \rho =1 \ , \label{10}\ee and therefore $\cD \rho = \d
\rho / \d t_0 \cD t_0  = P\cD t_0 =1 $, which yields \be \cD t_0 =
P^{-1} . \ee As a result, Eq.~(\ref{4}) takes the form \be \dot A =
-(\gamma P),_{\bar Y} , \label{11} \ee where $\dot {( \ )} \equiv
\d_{t_0}$.

For the stationary background  considered here, $P=2^{-1/2}(1+Y\bar
Y)$, and $\dot P = 0$.  The coordinates $Y$,  and $\t$ are
independent from $\bar Y$, which allows us to integrate
Eq.~(\ref{11}). We obtain the following general solution \be \gamma
= - P^{-1}\int \dot A d\bar Y =
 \frac{2^{1/2}\dot \psi} {P^2 Y} +\phi (Y,\t)/P ,
\label{12}\ee where $\phi$ is an arbitrary analytic function of $Y$
and $\t$. We set $\phi (Y,\t) =0.$ The term $\gamma$  in $ \cF _{31} =\gamma Z - (AZ),_1  \ $
describes a part of the null electromagnetic radiation   which falls
off asymptotically as $1/r$ and propagates along the Kerr principal
null congruence $e^3$. As it was discussed in \cite{Bur-nst}, it
describes a loss of mass by radiation with the stress-energy tensor
$\kappa T^{(\gamma)}_\mn = \frac 12 \gamma \bar \gamma e^3_{\m}
e^3_{\n}$.

We now evaluate the term $(AZ),_1. $
 For the stationary case we have the relations $Z,_1
=2ia \bar Y (Z/P)^3 $ and  $\t,_1 =- 2ia \bar Y Z/P^2 $ . This
yields \be (AZ),_1 = \frac{Z}{P^2} (\psi ,_Y - 2ia  \dot \psi
\frac{\bar Y}{P^2} - 2 \psi \frac{P_Y}{P}) + A 2ia \frac{Z \bar
Y}{P^3} . \label{AZ1} \ee Since $Z/P =1/(r+ia \cos \theta)$, this
expression contains the terms which fall off like $r^{-2}$ and
$r^{-3}$. However, it contains also the factors which depend on the
coordinate $Y = e^{i\phi} \tan \frac {\theta} 2 $ and can be
singular at the $z$-axis, forming the narrow beams, i.e. the
half-infinite lines of singularity. In particular, it can be the
$z^+$ or $z^-$ axis, which correspond to $\theta =0$ and
$\theta=\pi$ (cases $n=\pm1$, respectively).

\end{document}